\documentclass[12pt]{article}
\pdfoutput=1

\usepackage[utf8]{inputenc}
\usepackage[margin=1in]{geometry}
\usepackage[tbtags]{amsmath} 				
\usepackage{amssymb, mathrsfs}			
\usepackage{physics}
\usepackage{graphicx}			
\usepackage{tensor}				
\usepackage[makeroom]{cancel}	
\usepackage{bm}								
\usepackage[
      colorlinks=true,
      linkcolor=blue,
      urlcolor=blue,
      filecolor=black,
      citecolor=red,
      pdfstartview=FitV,
      pdftitle={},
      pdfauthor={Michael Gutperle},
      bookmarksopen=true
      ]{hyperref}
\usepackage{subcaption}

\usepackage{graphicx,calc}
\newlength\myheight
\newlength\mydepth
\settototalheight\myheight{Xygp}
\usepackage[utf8]{inputenc}

\def\Tr{{\rm Tr}}
\def\det{{\rm det \,}}

\renewcommand{\Im}{\operatorname{Im}}
\renewcommand{\Re}{\operatorname{Re}}
\DeclareMathOperator{\diag}{diag}

\def\cF{\mathcal{F}}
\def\cD{\mathcal{D}}

\def\cK{\mathcal{K}}
\def\cL{\mathcal{L}}

\def\cN{\mathcal{N}}
\def\cO{\mathcal{O}}

\def\cW{\mathcal{W}}

\def\SU{\text{SU}}
\def\U{\text{U}}
\def\SO{\text{SO}}
\def\AdS{\text{AdS}}
\def\OSp{\text{OSp}}

\DeclareMathOperator{\Vol}{Vol}

\usepackage{sectsty}
\sectionfont{\large}

\renewcommand{\thefootnote}{\fnsymbol{footnote}}
\renewcommand{\thanks}[1]{\footnote{#1}}
\newcommand{\starttext}{
\setcounter{footnote}{0}
\renewcommand{\thefootnote}{\arabic{footnote}}}

\usepackage{dsfont}

\numberwithin{equation}{section} 		

\long\def\symbolfootnote[#1]#2{\begingroup%
\def\thefootnote{\fnsymbol{footnote}}\footnote[#1]{#2}\endgroup}

\begin{document}
\setlength{\baselineskip}{16pt}

\starttext
\setcounter{footnote}{0}

\begin{flushright}
\today
\end{flushright}

\bigskip

\begin{center}

{\Large \bf  Holographic Line Defects in $D=4$, $N=2$ Gauged Supergravity}

\vskip 0.4in

{\large Kevin Chen, Michael Gutperle,  and Matteo Vicino}

\vskip 0.2in

{\sl Mani L. Bhaumik Institute for Theoretical Physics}\\
{\sl Department of Physics and Astronomy }\\
{\sl University of California, Los Angeles, CA 90095, USA} 

\bigskip

\end{center}
 
\begin{abstract}
\setlength{\baselineskip}{16pt}

We construct solutions of four-dimensional $N=2$ gauged supergravity coupled to vector multiplets which are holographically dual to superconformal line defects. 
For the gauged STU and the $\SU(1,n)/\U(1)\times \SU(n)$ coset models, we use the solutions to calculate holographic observables such as the expectation value of the defect and one-point functions in the presence of the defect.
 \end{abstract}

\setcounter{equation}{0}
\setcounter{footnote}{0}

\newpage

\section{Introduction}
\label{sec:intro}

Extended objects such as defects, line operators, and interfaces are important tools in the study of quantum field theories. 
For conformal field theories (CFTs) with holographic duals, in many cases the extended objects have a realization on the gravity side. 
One example  of such a duality  is the fundamental string in $\AdS_5\times S^5$ describing a Wilson line in the fundamental representation \cite{Maldacena:1998im,Rey:1998ik}, and its generalization to D3 probe branes with an $\AdS_2\times S^2$ worldvolume and D5 probe branes with an $\AdS_2\times S^4$ worldvolume \cite{Drukker:2005kx,Hartnoll:2006is,Yamaguchi:2006tq,Gomis:2006sb,Gomis:2006im}. 
When the number of probe branes becomes large, the backreaction can not be neglected and a fully backreacted  type IIB supergravity solution replaces the probe description. 
For the case of the half-BPS Wilson loop, this solution was found in \cite{DHoker:2007mci}.\footnote{See \cite{Lunin:2006xr} for earlier work on IIB supergravity solutions for Wilson loops.}
In general, the fully backreacted solutions in type II and M-theory are  warped product geometries and the solutions are complicated and difficult to obtain.\footnote{See   \cite{Gauntlett:2006qw,Gauntlett:2006ns} for  gravity solutions dual to conformal defects in both type IIB and
D=11 supergravity.}
A simpler setting is to consider lower-dimensional gauged supergravities, where the ansatz is simpler and the number of fields is smaller. 
If the gauged supergravity is a consistent truncation of a higher-dimensional theory,  the resulting lower-dimensional solution can be uplifted to produce solutions in the higher-dimensional theory.

In this paper, we consider four-dimensional $N=2$ gauged supergravity,  which has been used in the past  to describe condensed matter systems in three dimensions in order to find holographic models for superfluids and superconductors, see e.g.~\cite{Hartnoll:2008vx,Balasubramanian:2008dm,Gubser:2008wv}.  
We construct half-BPS supergravity solutions which are dual to line defects in three-dimensional $\cN=2$ superconformal field theories. 
The metric ansatz is given by $\AdS_2\times S^1$ warped over an interval. 
We generalize the analysis of \cite{Gutperle:2018fea}, which considered pure gauged supergravity, to the case of matter couplings. 

The structure of the paper is as follows. 
In section  \ref{sec:sugra}, we review our conventions for four-dimensional $N=2$ gauged supergravity coupled to vector multiplets. 
In section  \ref{sec:defect}, we give a general solution describing a half-BPS line defect, obtained by a double analytic continuation of the  black hole solutions first found by Sabra \cite{Sabra:1999ux}.
Since the behavior of the vector multiplet scalars can only be determined implicitly, we consider three examples, namely a single scalar model, the gauged STU model, and the $\SU(1,n)$ coset model to obtain explicit solutions. 
In section \ref{sec:holo}, we use the machinery of holographic renormalization  to calculate holographic observables for the solutions, namely the on-shell action and the expectation values of operators dual to the supergravity fields.
In section \ref{sec:conical}, we explore the conditions for a regular geometry and calculate their consequences.
In section \ref{sec:discussion}, we discuss our results and possible directions for future research.
Our conventions  and some details of the calculations presented in the main body of the paper are relegated to several  appendices.

\section{$D=4$, $N=2$ Gauged Supergravity}
\label{sec:sugra}

In this section, we review four-dimensional $N = 2$ gauged supergravity coupled to $n$ vector multiplets.
We use the conventions and notations of \cite{Lauria:2020ahi, Cacciatori:2008ek, Cacciatori:2009iz}.
 
The field content of the gauged supergravity theory is as follows.
The supergravity contains one graviton $e^a_\mu$, two gravitinos $\psi^i_\mu$, and one graviphoton $A^0_\mu$.
The gravity multiplet can be coupled to $N=2$ matter, and in particular we consider $n$ vector multiplets, which are labeled by an index $\alpha = 1, 2, \dotsc, n$.
Each vector multiplet contains one vector field $A^\alpha_\mu$, two gauginos $\lambda^\alpha_i$, and one complex scalar $\tau^\alpha$. 
In this paper we do not consider adding $N=2$ hypermultiplets.

It is convenient to introduce a new index $I = 0, 1, \dotsc, n$ and include the graviphoton with the other vector fields as $A^I_\mu$.
The complex scalars $\tau^\alpha$ parametrize a special K\"ahler manifold equipped with a holomorphic symplectic vector
\begin{align} 
v(\tau) = \mqty( Z^I(\tau) \\ \cF_I(\tau) )~,
\end{align}
where the K\"ahler potential $\cK(\tau, \bar\tau)$ is determined by
\begin{align}
e^{- \cK(\tau, \bar\tau)} = -i \ev{v, \bar{v}} \equiv -i(Z^I \bar{\cF}_I - \cF_I \bar{Z}^I)~.
\end{align}
In the models we will consider, there exists a holomorphic function $\cF(Z)$, called the prepotential, that is homogeneous of second order in $Z$ such that
\begin{align}
\cF_I(\tau) = \pdv{Z^I} \cF\qty\big(Z(\tau))~.
\end{align}
The supergravity theory is fully specified by the prepotential $\cF(Z)$ and the choice of gauging of the $\SU(2)$ $R$-symmetry. 
We will choose the $\U(1)$ Fayet-Iliopoulos (FI) gauging.
The only charged fields of the theory are the gravitinos, which couple to the gauge fields through the linear combination $\xi_I A^I$, for some real constants $\xi_I$.
The two gravitinos have opposite charges $\pm g \xi_I$ for each $\U(1)$ gauge factor, where $g$ is the gauge coupling.

The bosonic action is\footnote{We set $8\pi G_{\rm N} = 1$.} 
\begin{align}
e^{-1} \cL_{\rm bos} &= \frac{1}{2} R - g_{\alpha \bar \beta} \partial^\mu \tau^\alpha \partial_\mu {\bar \tau}^{\bar \beta} - V(\tau, \bar\tau) \nonumber \\
&\phantom{=} + \frac{1}{4} (\Im \cN)_{IJ} F^{I\mu\nu} F^J_{\mu\nu} - \frac{1}{8} (\Re \cN)_{IJ} e^{-1} \epsilon^{\mu\nu\rho\sigma} F^I_{\mu\nu} F^J_{\rho\sigma}~, \label{eq:sugra-action}
\end{align}
where $F^I_{\mu\nu} = \partial_\mu A^I_\nu - \partial_\nu A^I_\mu$ are the field strengths and $g_{\alpha \bar \beta} = \partial_\alpha \partial_{\bar \beta} \cK$ is the K\"ahler metric of the scalar manifold.
We use $G_{\mu\nu}$ to denote the four-dimensional metric, so $e = \sqrt{- \det G}$.
The scalar potential is
\begin{align}
V(\tau, \bar\tau) = -2 g^2 \xi_I \xi_J \qty((\Im \cN)^{-1|IJ} + 8 e^{\cK} \bar{Z}^I Z^J)~,
\end{align}
where the kinetic matrix $\cN_{IJ}$ is given by
\begin{align}
\cN_{IJ}(\tau, \bar\tau) &= \bar{\cF}_{IJ} + 2i \frac{(\Im \cF_{IL})( \Im \cF_{JK}) Z^L Z^K}{(\Im \cF_{MN}) Z^M Z^N}~, & \cF_{IJ} &\equiv \pdv{Z^I} \pdv{Z^J} \cF(Z)~.
\end{align}
This is equivalently defined as the matrix which solves the equations
\begin{align}
\cF_I &= \cN_{IJ} Z^J~, & \cD_{\bar\alpha} \bar{\cF}_I &= \cN_{IJ} \cD_{\bar\alpha} \bar{Z}^J~,
\end{align}
where $\cD$ is the K\"ahler covariant derivative,
\begin{align}
\cD_\alpha v &= (\partial_\alpha + \partial_\alpha \cK) v~, \nonumber \\
\cD_{\bar\alpha} \bar{v} &= (\partial_{\bar\alpha} + \partial_{\bar\alpha} \cK) \bar{v}~, \nonumber \\
\cD_\alpha \bar{v} &= \partial_\alpha \bar{v} = 0~, \nonumber \\
\cD_{\bar\alpha} v &= \partial_{\bar\alpha} v = 0~. 
\end{align}
The equations of motion are obtained by varying the Lagrangian \eqref{eq:sugra-action} 
\begin{align}
R_{\mu\nu} &= 2 g_{\alpha \bar\beta} \partial_\mu \tau^\alpha \partial_\nu {\bar\tau}^{\bar\beta} +  V G_{\mu\nu}  + (\Im \cN)_{IJ} \qty( - \tensor{F}{^I_\mu^\rho} \tensor{F}{^J_{\nu\rho}} + \frac{1}{4} F^{I\rho\sigma} \tensor{F}{^J_{\rho\sigma}} G_{\mu\nu} )~, \nonumber \\
\partial_\mu \qty( e g_{\alpha \bar \beta} \partial^\mu {\bar\tau}^{\bar\beta} ) &= e \qty( (\partial_\alpha g_{\beta\bar\gamma}) \partial^\mu \tau^\beta \partial_\mu {\bar\tau}^{\bar\gamma} - \frac{1}{4} \partial_\alpha (\Im \cN)_{IJ} F^{I\mu\nu} F^J_{\mu\nu} + \partial_\alpha V ) \nonumber \\
&\phantom{=} + \frac{1}{8} \partial_\alpha (\Re \cN)_{IJ} \epsilon^{\mu\nu\rho\sigma} F^I_{\mu\nu} F^J_{\rho\sigma}~, \nonumber \\
0 &= \partial_\mu \qty( e (\Im \cN)_{IJ} F^{J\mu\nu} -  \frac{1}{2} (\Re \cN)_{IJ} \epsilon^{\mu\nu\rho\sigma} F^J_{\rho\sigma} )~. \label{eq:sugra-eom}
\end{align}
The supersymmetry transformations are given in appendix \ref{sec:susy}.

\section{Line Defect Solution}
\label{sec:defect}

In this section, we give a general solution describing a half-BPS line defect in four-dimensional $N=2$ gauged supergravity, and then construct the solution for three specific choices of the prepotential.

\subsection{Holographic Line Defects}
\label{sec:defect-holo}

A conformal line defect in three dimensions is a codimension-two defect which breaks the three-dimensional conformal group $\SO(3,2)$ down to an $ \SO(2,1) \times \SO(2)$ subgroup. 
The subgroup factors represent the unbroken conformal symmetry along the defect and transverse rotations about the defect, respectively.
Minkoswski space  $\mathbb{R}^{1,2}$ is related by a Weyl transformation to $\AdS_2\times S^1$, namely
\begin{align}
-\dd{t^2}+ \dd{r^2} + r^2 \dd{\phi^2} = \Omega(r) \left( {-\dd{t^2}+ \dd{r^2} \over r^2}+ \dd{\phi^2} \right)~.
\end{align}
Hence in the holographic dual,  the $ \SO(2,1) \times \SO(2)$ symmetry can be realized as the isometries of $\AdS_{2} \times S^{1}$, which we choose as the  boundary of the four-dimensional asymptotically anti-de Sitter space. 
Therefore we consider a metric ansatz with  $\AdS_{2} \times S^{1}$ warped over a radial coordinate. We note that the location of the defect at $r=0$ in Minkowski space gets mapped to the boundary of $\AdS_2$ in the $\AdS_2\times S^1$ geometry. Secondly, the absence of a conical singularity on the boundary fixes the periodicity of the angle $\phi$ to be $2\pi$.

The superconformal algebras in three dimensions are $\OSp({\cal N}|4)$, where ${\cal N}= 1,2,\dotsc,6,8$. 
For the CFT dual of four-dimensional $N=2$ gauged supergravity, the relevant superalgebra is $\OSp(2|4)$ which has four Poincar\'e and four conformal supercharges. 
A conformal line defect is called superconformal if it preserves some supersymmetry. 
In the present paper, we will consider half-BPS defects which preserve an $\OSp(2|2)$ superalgebra and hence four  of the eight supersymmetries.

\subsection{General Solution}
\label{sec:defect-general}

Four-dimensional $N=2$, $\U(1)$ FI gauged supergravity admits half-BPS black hole solutions first found in \cite{Sabra:1999ux}. 
The line defect solutions  with $\AdS_2 \times S^1$ geometry are constructed  by a double analytic continuation of the black hole solution.
The metric and gauge fields are given by
\begin{align}
\dd{s^2} &= r^2 \sqrt{H(r)} \dd{s^2_{\AdS_2}} + \frac{f(r)}{\sqrt{H(r)}}  \dd{s^2_{S^1}} + \frac{\sqrt{H(r)}}{f(r)} \dd{r^2}~, \nonumber \\
f(r) &= -1 + 8 g^2 r^2 H(r)~, \nonumber \\
H(r)^{1/4} &= \frac{1}{\sqrt{2}} e^{\cK/2} Z^I H_I(r)~,\nonumber\\
H_I(r) &= \xi_I + \frac{q_I}{r}~, \quad \quad I=0,1,\dotsc,n~, \nonumber \\
A^I &= \qty( -2 H(r)^{-1/4} e^{\cK/2} Z^I + \mu^I ) \dd{\theta}~, \quad \quad I=0,1,\dotsc,n~, \label{eq:sabra}
\end{align}
for some real constants $q_I$ and $\mu_I$, where $Z^I = \bar{Z}^I$. 
Given a prepotential $\cF(Z)$ and choice of parametrization of the symplectic sections $Z^I(\tau)$, the scalars $\tau^\alpha$ are given implicitly by the equation
\begin{align}\label{eq:sabrascal}
i H^{1/4} e^{\cK/2} (\cF_I - \bar{\cF}_I) = \frac{1}{\sqrt{2}} H_I~.
\end{align} 
At the conformal boundary where $r \to \infty$, in order to have asymptotic $\AdS_4$ we need $2\sqrt{2} g \theta$ to be $2\pi$-periodic, i.e.~$\theta \sim \theta + \pi/\sqrt{2}g$.
The $\AdS_4$ length scale is then given by
\begin{align}
L^{-2} = 8 g^2 H(r = \infty)^{1/2}~. \label{eq:ads-length}
\end{align}
We will set $8g^2 = 1$ to obtain the usual $S^1$ periodicity $\theta \sim \theta + 2\pi$.

The center of the space\footnote{For the black hole geometry this is the location of the horizon.} $r = r_+$ corresponds to the largest value of $r$ where $f(r) = 0$.
We consider radii taking values in the range $r \in [r_+, \infty)$. 
Demanding a regular geometry also requires $r_+ > 0$ and the absence of a conical singularity at the center of the space, both of which can be done by tuning the $q_I$ and $\xi_I$ parameters.
This is explored in further detail in section \ref{sec:conical}.

\subsection{Examples }
\label{sec:defect-ex}

For a general prepotential, the equation (\ref{eq:sabrascal}) is very complicated and can only be solved numerically. 
Consequently, we will explicitly work out the line defect solution for three specific prepotentials, for which we can find explicit expressions for the scalars.
An important requirement is the existence of an $\AdS_4$ vacuum, which not all prepotentials admit, see e.g.  \cite{Cacciatori:2009iz, Hristov:2010ri}.

\subsubsection{Single Scalar Model}
\label{sec:defect-ex-single}

Consider a single ($n=1$) vector multiplet with the prepotential $\cF(Z) = -i Z^0 Z^1$.
This theory has a single complex scalar $\tau$ and the scalar manifold is $\smash{\frac{\SU(1, 1)}{\U(1)}}$.
Using the parametrization $(Z^0, Z^1) = (1, \tau)$, we can calculate the K\"ahler potential, kinetic matrix, and scalar potential,
\begin{align}
e^{\cK(\tau, \bar\tau)} &= \frac{1}{2 (\tau + \bar\tau)}~, \nonumber \\
\cN(\tau, \bar\tau) &= -i \mqty( \tau & 0 \\ 0 & 1/\tau ) ~, \nonumber \\
V(\tau, \bar\tau) &= -\frac{1}{2(\tau + \bar\tau)} \qty( \xi_0^2 + 2 \xi_0 \xi_1 (\tau + \bar\tau) + \xi_1^2 \tau \bar\tau )~.
\end{align}
The potential has extrema at $\tau = \pm \xi_0/\xi_1$, but only $\tau=\xi_0/\xi_1$ maintains $e^{\mathcal{K}} >0$ for $\xi_I > 0$. The cosmological constant at this extremum gives the $\AdS_{4}$ length scale
\begin{equation}
L^{-2} = \frac{1}{2} \xi_0 \xi_1.
\end{equation}
We choose $\xi_1 = 2/\xi_0$ to set the $\AdS_{4}$ length scale to unity. 
The line defect solution \eqref{eq:sabra}  has the explicit form,
\begin{align}
\dd{s^2} &= r^2 \sqrt{H} \dd{s^2_{\AdS_2}} + \frac{f}{\sqrt{H}} \dd{s^2_{S^1}} + \frac{\sqrt{H}}{f} \dd{r^2}~, \nonumber \\
f(r) &= -1 + r^2 H(r)~, \nonumber \\
\sqrt{H(r)} &= \frac{1}{2} H_0 H_1~, \nonumber\\
 H_I(r) &= \xi_I + \frac{q_I}{r}~,\quad \quad I=0,1~, \nonumber \\
A^I &= \qty(  - \frac{\sqrt{2}}{H_I} + \mu^I) \dd\theta~,\quad \quad I=0,1~.\label{eq:defect-single-scalar1}
\end{align}
The scalar is given by
\begin{equation}
\tau = \frac{H_0}{H_1}~. \label{eq:defect-single-scalar2}
\end{equation}
We have verified that the above fields obey the equations of motion \eqref{eq:sugra-eom}.

\subsubsection{Gauged STU Model}
\label{sec:defect-ex-stu}

The STU model is given by considering $n=3$ vector multiplets with the prepotential 
\begin{align}
\cF(Z) &= -2i \sqrt{Z^0 Z^1 Z^2 Z^3}.
\end{align}
This theory has three complex scalars $\tau^1, \tau^2, \tau^3$ and the scalar manifold is three copies of $\smash{\frac{\SU(1, 1)}{\U(1)}}$. 
When all $\xi_I = \xi > 0$ are equal, this theory is a consistent truncation of $\cN=8$ gauged supergravity \cite{Duff:1999gh, Cvetic:1999xp}.
For reference on this model, see \cite{Cabo-Bizet:2017xdr}.
Using the parametrization $(Z^0, Z^1, Z^2, Z^3) = (1, \tau^2 \tau^3, \tau^1 \tau^3, \tau^1 \tau^2)$, the K\"ahler potential is
\begin{align}
e^{\cK(\tau, \bar\tau)} &= \frac{1}{(\tau^1 + {\bar\tau}^1)(\tau^2 + {\bar\tau}^2)(\tau^3 + {\bar\tau}^3)}~.
\end{align}
The expressions for the kinetic matrix and scalar potential are complicated, but simplify for real scalars $\tau^\alpha = \bar{\tau}^{\bar\alpha}$, which will be the case for the line defect solution.
\begin{align}
\cN(\tau, \bar{\tau} = \tau) &= -i \diag \qty(\tau^1\tau^2\tau^3, \frac{\tau^1}{\tau^2\tau^3}, \frac{\tau^2}{\tau^1\tau^3}, \frac{\tau^3}{\tau^1\tau^2} )~, \nonumber \\
V(\tau, \bar{\tau} = \tau) &= - \frac{1}{2} \qty( \xi_0 \qty( \frac{\xi_1}{\tau^1} + \frac{\xi_2}{\tau^2} + \frac{\xi_3}{\tau^3}) + \qty( \tau^1 \xi_2 \xi_3 + \xi_1 \tau^2 \xi_3 + \xi_1 \xi_2 \tau^3) )~.
\end{align}
The potential has extrema at
\begin{align}
\tau^1 &= \pm \sqrt{\frac{\xi_0 \xi_1}{\xi_2 \xi_3}}~, & \tau^2 &= \pm \sqrt{\frac{\xi_0 \xi_2}{\xi_1 \xi_3}}~, & \tau^3 &= \pm \sqrt{\frac{\xi_0 \xi_3}{\xi_1 \xi_2}}~.
\end{align}
Positivity of $e^{\mathcal{K}}$ requires us to choose the positive root. The cosmological constant at this extremum gives the $\AdS_4$ length scale
\begin{align}
L^{-2} =  \sqrt{\xi_0 \xi_1 \xi_2 \xi_3}~.
\end{align}
We pick the non-zero constants $\xi_I$ in a way that sets the $\AdS_4$ length scale to unity.
The line defect solution \eqref{eq:sabra} has the explicit form,
\begin{align}
\dd{s^2} &= r^2 \sqrt{H} \dd{s^2_{\AdS_2}} + \frac{f}{\sqrt{H}} \dd{s^2_{S^1}} + \frac{\sqrt{H}}{f} \dd{r^2}~, \nonumber \\
f(r) &= -1 + r^2 H(r)~, \nonumber \\
H(r) &= H_0 H_1 H_2 H_3~,\nonumber\\
 H_I(r) &= \xi_I + \frac{q_I}{r}~, \quad \quad I=0,1,2,3~, \nonumber \\
A^I &= \qty( -\frac{1}{\sqrt{2} H_I}+ \mu^I ) \dd{\theta}~, \quad \quad I=0,1,2,3~. \label{eq:defect-stu1}
\end{align}
The scalars are
\begin{align}
\tau^1 &= \sqrt{\frac{H_0 H_1}{H_2 H_3}}~, & \tau^2 &= \sqrt{\frac{H_0 H_2}{H_1 H_3}}~, & \tau^3 &= \sqrt{\frac{H_0 H_3}{H_1 H_2}}~. \label{eq:defect-stu2}
\end{align}
This solution is also the double analytic continuation of the hyperbolic black hole solution in \cite{Hosseini:2019and}.
As consistency checks, we have verified that the above solution obeys the equations of motion \eqref{eq:sugra-eom} and is half-BPS.
The latter was done by a direct calculation, independent of \cite{Sabra:1999ux}, which can be found in appendix \ref{sec:susy}.
\subsubsection{$\SU(1, n)$ Coset Model}
\label{sec:defect-ex-coset}

Another model which admits an $\AdS_4$ vacuum has the prepotential $\cF(Z) = \frac{i}{4} Z^I \eta_{IJ} Z^J$, and can be formulated with any number of vector multiplets.
$\eta_{IJ}$ is a Minkowski metric, which we will take to be $\eta = \diag(-1, +1, \dotsc, +1)$.
The scalar manifold of this theory is $\smash{\frac{\SU(1, n)}{\U(1) \times \SU(n)}}$.
Using the parametrization $(Z^0, Z^\alpha) = (1, \tau^\alpha)$, the K\"ahler potential is
\begin{align}
e^{\cK(\tau, \bar\tau)} &= \frac{1}{1 - \sum_{\alpha} \tau^{\alpha} \bar{\tau}^{\alpha}}~.
\end{align}
Once again, the kinetic matrix and scalar potential have simpler forms for real scalars $\tau^\alpha = \bar{\tau}^{\bar\alpha}$.
The matrix $\eta_{IJ}$ is used to lower indices, e.g.~$Z_I = \eta_{IJ} Z^J$.
\begin{align}
\cN_{IJ}(\tau, \bar{\tau} = \tau) &= - \frac{i}{2}\eta_{IJ} - i e^{\cK(\tau, \tau)} Z_I Z_J~, \nonumber \\
V(\tau, \bar{\tau} = \tau) &= \frac{1}{2} \xi_I \eta^{IJ} \xi_J  - \frac{(\xi_0 + \sum_{\alpha} \xi_\alpha \tau^\alpha)^2}{1 - \sum_{\alpha}(\tau^\alpha)^2}~.
\end{align}
This potential has an extremum at $\tau^\alpha = - \xi_\alpha / \xi_0$.\footnote{The other extrema at $\xi_0 + \sum_\alpha \xi_\alpha \tau^\alpha = 0$ do not admit $\AdS_4$ vacua while maintaining $e^\cK$ positive.}
The cosmological constant at this extremum gives us the $\AdS_4$ length scale
\begin{align}
L^{-2} = -\xi^2 / 2~,
\end{align}
where $\xi^2 = \xi_I \eta^{IJ} \xi_J$. 
We pick a time-like $\xi_I$ with $\xi^2 = -2$ that will set the $\AdS_4$ length scale to unity.
The line defect solution \eqref{eq:sabra} has the explicit form,
\begin{align}
\dd{s^2} &= r^2 \sqrt{H} \dd{s^2_{\AdS_2}} + \frac{f}{\sqrt{H}} \dd{s^2_{S^1}} + \frac{\sqrt{H}}{f} \dd{r^2}~, \nonumber \\
f(r) &= -1 + r^2 H(r)~, \nonumber \\
\sqrt{H(r)} &= - \frac{1}{2} H_I \eta^{IJ} H_J~,\nonumber\\
 H_I(r) &= \xi_I + \frac{q_I}{r}~, \quad \quad I=0,1,\dotsc,n~, \nonumber \\
A^I &= \qty( \frac{\sqrt{2}\eta^{IJ} H_J}{\sqrt{H}}  + \mu^{I} ) \dd{\theta}~, \quad \quad I=0,1,\dotsc,n~.
\label{eq:defect-coset-1}
\end{align}
The scalars are
\begin{align}
\tau^\alpha = -\frac{H_\alpha}{H_0}~.
\label{eq:defect-coset-2}
\end{align}
We have verified that the above fields obey the equations of motion \eqref{eq:sugra-eom}.

\section{Holographic Calculations}
\label{sec:holo}

In this section, we use the machinery of holographic renormalization \cite{deHaro:2000vlm, Skenderis:2002wp} to calculate the on-shell action and the  one-point functions of dual operators of the boundary CFT in the presence of the defect, namely the stress tensor, scalar, and currents. 
This is done explicitly for the three examples in section \ref{sec:defect-ex}.

\subsection{General Procedure}
\label{sec:holo-general}

First, we put the metric into the Fefferman-Graham (FG) form,
\begin{align} 
\dd{s^2} = \frac{1}{z^2} \qty( \dd{z^2} + g_{ij}(x, z) \dd{x^i} \dd{x^j} )~,
\end{align}
where $i,j = 1, 2, 3$ run over the $\AdS_2$ and $S^1$ indices and $z \to 0$ is the conformal boundary.
This is done by taking $z = z(r)$ so that the appropriate coordinate change is obtained by the solution to the ordinary differential equation
\begin{align}
-\frac{H(r)^{1/4}}{f(r)^{1/2}} \dd{r} = \frac{\dd{z}}{z}~, \label{eq:fg-ode}
\end{align}
which can be integrated perturbatively in $1/r$.
This coordinate change gives the FG expansions of the fields, which we assume will take the form
\begin{align}
g_{ij} &= {g_0}_{ij} + z^2 {g_2}_{ij} + z^3 {g_3}_{ij} + \order{z^4}~, \nonumber \\
A^I &= A^I_0 + z A^I_1 + \order{z^2}~, \nonumber \\
\tau^\alpha &= \tau^\alpha_0 + z \tau^\alpha_1 + z^2 \tau^\alpha_2  + \order{z^3}~, \nonumber \\
\bar{\tau}^{\bar\alpha} &= \tau^\alpha_0 + z \tau^\alpha_1 + z^2 \tau^\alpha_2  + \order{z^3}~,
\end{align}
where $A^I_0$ and $A^I_1$ are 1-forms on the $x^1, x^2, x^3$ coordinates.
The constants $\tau^\alpha_0$ are the $\AdS_4$ vacuum values of the scalars, which depend on the model. There is no gravitational conformal anomaly (i.e.~a term proportional to $z^3 \log z$ in the expansion of $g_{ij}$) since $d = 3$ is odd. \\

In the three-dimensional boundary CFT, the conformal dimensions of the dual operators  corresponding to the scalars $\tau^\alpha$ and vector fields $A^I$ are determined by the linearized bulk equations of motion near the AdS boundary. 
For instance, using the expansion $\tau^\alpha \sim \tau^\alpha_0 + z^{\Delta_\tau}$ in the linearized equation of motion for the scalar, we find that the scaling dimension of the dual operator is related to the mass-squared of the field by the equation
\begin{align}
\Delta_\tau (\Delta_\tau - 3) = -2~.
\end{align} 
The mass-squared is $-2$ for all scalars of the three examples considered in this paper.
This mass-squared is within the window where both standard and alternative quantization are possible \cite{Klebanov:1999tb}, which implies that the scaling dimension of the dual operator can be either $\Delta_\tau = 1$ or $\Delta_\tau = 2$.
Similarly, using the expansion $A^I \sim z^{\Delta_A - 1} \dd{\theta}$ in the linearized equation of motion for the vector field gives us
\begin{align}
(\Delta_A - 1)(\Delta_A - 2) = 0~.
\end{align}
We must have $\Delta_A = 2$ as the vector field sources a conserved current of the boundary CFT.

These scaling dimensions naturally fit into the flavor current multiplet $A_2 \overline{A}_2[0]_1^{(0)}$ of the $d=3$, $\cN = 2$ boundary CFT, using the notation of \cite{Cordova:2016emh}.
This short multiplet contains, in addition to the spin-1 operator $[2]_2^{(0)}$ with scaling dimension $\Delta = 2$, two scalar operators $[0]_1^{(0)}$ and $[0]_2^{(0)}$ as bottom and top components with scaling dimensions $\Delta =1$ and $2$ respectively.
The stress tensor multiplet $A_1 \overline{A}_1[2]_2^{(0)}$ is also present, as usual.  \\

In the four-dimensional gauged supergravity, for a well-defined variational principle of the metric we need to add to the bulk action given by the Lagrangian \eqref{eq:sugra-action} the Gibbons-Hawking boundary term,
\begin{align}
I_{\rm bulk} &= \int_M \dd[4]{x}  \mathcal{L}_{\rm bos}~, \nonumber \\
I_{\rm GH} &= \int_{\partial M} \dd[3]{x} \sqrt{-h} \, \Tr(h^{-1} K)~,
\end{align}
where $h_{ij}$ is the induced metric on the boundary and $K_{ij}$ is the extrinsic curvature.
In FG coordinates, these take the form,
\begin{align}
h_{ij} &= \frac{1}{z^2} g_{ij}~, & K_{ij} &= -\frac{z}{2} \partial_z h_{ij}~.
\end{align}
The action $I_{\rm bulk} + I_{\rm GH}$ diverges due to the infinite volume of integration.
To regulate the theory, we restrict the bulk integral to the region $z \geq \epsilon$ and evaluate the boundary term at $z = \epsilon$.
Divergences in the action then appear as $1/\epsilon^k$ poles.\footnote{In even boundary dimensions, a term proportional to $\log \epsilon$ may also appear.}
Counterterms $I_{\rm ct}$ are added on the boundary which subtract these divergent terms.
The counterterms have been constructed in \cite{Cabo-Bizet:2017xdr} and are compatible with supersymmetry. They are
\begin{align}
I_{\rm ct} &= \int_{\partial M} \dd[3]{x} \sqrt{-h} \, \qty( \cW -\frac{1}{2} R[h] )~, & \cW &\equiv - \sqrt{2} e^{\cK/2} \qty| \xi_I Z^I|~,
\end{align}
where $R[h]$ is the Ricci scalar of the boundary metric and $\mathcal{W}$ is the superpotential.
In all, the renormalized action,
\begin{align}
I_{\rm ren} = I_{\rm bulk} + I_{\rm GH}+ I_{\rm ct} \label{eq:action-ren}~,
\end{align}
is finite. 
We can then take functional derivatives to obtain finite expectation values of the dual CFT operators. 
Let $T_{ij}$ be the boundary stress tensor, $\cO_\alpha$ be the operators dual to $\tau^\alpha$, and ${J_I}_i$ be the current operators dual to $A^I_\mu$.

\subsubsection{Stress Tensor Expectation Value}
\label{sec:holo-general-stress}

The expectation value of the boundary stress tensor is defined to be \cite{Balasubramanian:1999re}
\begin{align}
\ev{T_{ij}} \equiv \frac{-2}{\sqrt{-g_0}} \fdv{ I_{\rm ren}}{g_0^{ij}}~.
\end{align}
The variation decomposes into two contributions: one coming from the regularized action and one coming from the counterterms.
As usual \cite{Kraus:2006wn}, the former is given by
\begin{align}
T^{\rm reg}_{ij}[h] \equiv \frac{-2}{\sqrt{-h}} \fdv{ (I_{\rm bulk} + I_{\rm GH}) }{{h}^{ij}} = - K_{ij} + h_{ij} \Tr(h^{-1} K)~.
\end{align}
The latter is straightforward to compute, and is given by
\begin{align}
T^{\rm ct}_{ij}[h] \equiv \frac{-2}{\sqrt{-h}} \fdv{ I_{\rm ct}}{{h}^{ij}} = h_{ij} \qty( \cW - \frac{1}{2} R[h]) + R_{ij}[h]~.
\end{align}
Therefore, 
\begin{align}
\ev{T_{ij}} = \lim_{\epsilon \to 0} \qty[\epsilon^{-1} \eval{\qty\Big( T^{\rm reg}_{ij}[h] + T^{\rm ct}_{ij}[h])}_{z = \epsilon}]~. \label{eq:stress-tensor-one-point}
\end{align}
By construction of the counterterms, this limit exists.

\subsubsection{Scalar Expectation Values}
\label{sec:holo-general-scalar}

The expectation value of the operator $\cO_\alpha$ is similarly defined by
\begin{align}
\ev{\cO_\alpha} \equiv \frac{1}{\sqrt{-g_0}} \fdv{ I_{\rm ren}}{\tau_1^\alpha}  = \lim_{\epsilon \to 0} \qty[ \epsilon^{-2} \eval{\frac{1}{\sqrt{-h}} \fdv{I_{\rm ren}}{\tau^\alpha}}_{z = \epsilon} ]~.
\label{eq:scalar-one-point}
\end{align}
The variation has contributions from the bulk action and the counterterms, and is
\begin{align}
\frac{1}{\sqrt{-h}} \fdv{I_{\rm ren}}{\tau^\alpha} = g_{\alpha\bar\beta} z \partial_z {\bar\tau}^{\bar\beta} + \partial_\alpha \cW~.
\end{align}
For real scalars, supersymmetry implies $\ev{\cO_\alpha} = 0$. A proof of this statement can be found in appendix \ref{sec:scalar-one-point}.
\subsubsection{Current Expectation Values}
\label{sec:holo-general-current}

The expectation value of the current operator $J_I$ is defined by
\begin{align}
\ev{J_I^i} \equiv \frac{1}{\sqrt{-g_0}} \fdv{ I_{\rm ren}}{{A^I_0}_i}  = \lim_{\epsilon \to 0} \qty[ \epsilon^{-3} \eval{\frac{1}{\sqrt{-h}} \fdv{I_{\rm ren}}{A^I_i}}_{z = \epsilon} ]~.
\label{eq:vector-one-point}
\end{align}
The only contribution to the variation comes from the bulk action, and is
\begin{align}
\frac{1}{\sqrt{-h}} \fdv{I_{\rm ren}}{A^I_i} = - (\Im \cN)_{IJ} h^{ij} z \partial_z A^J_j~.
\end{align}

\subsubsection{On-Shell Action}
\label{sec:holo-general-action}

We can evaluate the on-shell action for the line defect solution by further simplifying the bulk action to a total derivative \cite{Batrachenko:2004fd},
\begin{align}
\eval{I_{\rm bulk}}_{\text{ on-shell}} = \text{Vol}(\AdS_2) \text{Vol}(S^1) \eval{ \qty[ - \frac{H'(r)}{4H(r)} r^2 f(r) - r \qty\big(f(r) + 1) ] }^\infty_{r_+}~, \label{eq:r+action}
\end{align}
where $\Vol(S^1) = 2\pi$ and $\Vol(\AdS_2) = -2\pi$ is the regularized volume of $\AdS_2$.

\subsection{Examples}
\label{sec:holo-ex}
In this section, we use the general expressions derived in section \ref{sec:holo-general} to compute observables for the three examples considered in this paper.

\subsubsection{Single Scalar Model}
Let us consider the defect solution (\ref{eq:defect-single-scalar1}, \ref{eq:defect-single-scalar2}) for the single scalar model. 
The FG expansion of the radial coordinate $r$  from solving the ordinary differential equation \eqref{eq:fg-ode} is
\begin{equation}
\begin{split}
\frac{1}{r} = z &+ \frac{1}{2} \left(\sum_{I=0}^{1} \frac{q_I}{\xi_I} \right) z^2 + \frac{-16 + (3 q_1 \xi_0 + q_0 \xi_1) (3 q_0 \xi_1 + q_1 \xi_0)}{64} z^3 \\
& + \frac{(q_1 \xi_0 + q_0 \xi_1) (-16 + 12 q_0 q_1 \xi_0 \xi_1 + 3 (q_0 \xi_1+q_1 \xi_0)^2)}{384} z^4 + \order{z^5}~.
\end{split}
\end{equation} 
Using this coordinate change, the metric, gauge fields, and scalar can be expanded in FG coordinates.
The one-point functions in the presence of the line defect can then be evaluated by computing the limits (\ref{eq:stress-tensor-one-point}, \ref{eq:scalar-one-point}, \ref{eq:vector-one-point}) directly.
For the renormalized on-shell action \eqref{eq:action-ren}, the finite terms at the conformal boundary cancel, leaving just the term obtained by evaluating \eqref{eq:r+action} at $r = r_+$. 
In the end, we obtain the following expectation values:
\begin{align}
I_{\rm ren} &= \text{Vol}(\AdS_2) \text{Vol}(S^1) r_+~, \nonumber \\
\ev{T_{ij}} &= \frac{1}{2} \left( \sum_{I=0}^{1} \frac{q_I}{\xi_I} \right) \mqty(- g_{\AdS_2} & 0 \\ 0 & 2g_{S^1} )_{ij}~, \nonumber \\
\ev{T^i_i} &= 0~, \nonumber \\
\ev{\cO} &= 0~, \nonumber \\
\ev{{J_I}_i} &= \frac{q_I}{\sqrt{2}} \delta_{i\theta}~.\label{eq:single-evs}
\end{align}

\subsubsection{Gauged STU Model}
\label{sec:holo-ex-stu}

Let us consider the defect solution (\ref{eq:defect-stu1}, \ref{eq:defect-stu2}) for the gauged STU model.
Some of the calculations for this model are identical to those found in \cite{Hosseini:2019and}.
The FG expansion of the radial coordinate $r$ from solving the ODE \eqref{eq:fg-ode} is
\begin{align}
\frac{1}{r} = z + \frac{A}{4} z^2 + \frac{-16 + B_1 + 10 B_2}{64} z^3 + \frac{-16A + C_1 + 11 C_2 + 62 C_3}{384} z^4 + \order{z^5}~,
\end{align}
where we have defined the constants
\begin{align}
A &= \sum_{I=0}^3 \frac{q_I}{\xi_I}~, & B_1 &= \sum_{I=0}^3 \left(\frac{q_I}{\xi_I} \right)^2~, & B_2 &= \sum_{I < J} \frac{q_I q_J}{\xi_I \xi_J}~, \nonumber \\
C_1 &= \sum_{I=0}^3 \left(\frac{q_I}{\xi_I} \right)^3~, & C_2 &= \sum_{I \neq J} \left(\frac{q_I}{\xi_I} \right)^2 \frac{q_J}{\xi_J}~, & C_3 &= \sum_{I < J < K} \frac{q_I q_J q_K}{\xi_I \xi_J \xi_K}~.
\end{align}
Using this coordinate change, the fields of the defect solution can be expanded in FG coordinates.
We obtain the following on-shell action and one-point functions,
\begin{align}
I_{\rm ren} &= \text{Vol}(\AdS_2) \text{Vol}(S^1) r_+~, \nonumber \\
\ev{T_{ij}} &= \frac{1}{4} \left( \sum_{I=0}^{3} \frac{q_I}{\xi_I} \right) \mqty(- g_{\AdS_2} & 0 \\ 0 & 2g_{S^1} )_{ij}~, \nonumber \\
\ev{T^i_i} &= 0~, \nonumber \\
\ev{\cO_1} = \ev{\cO_2} = \ev{\cO_3} &= 0~, \nonumber \\
\ev{{J_I}_i} &= \frac{q_I}{\sqrt{2}} \delta_{i\theta}~.
\end{align}
Note that the expression for $I_{\text{ren}}$ is identical to that of the single scalar model, but the radius $r_{+}=r_{+}\left(\xi_{I},q_{I} \right)$ will be different.

\subsubsection{$\SU(1, n)$ Coset Model}
\label{sec:holo-ex-coset}
For the defect solution (\ref{eq:defect-coset-1}, \ref{eq:defect-coset-2}) of the $\SU(1, n)$ coset model, the FG expansion of the radial coordinate $r$ is
\begin{align}
\frac{1}{r} = z &- \frac{1}{2} q_{I} \xi^{I} z^2 - \frac{1}{4} \left[1 + \frac{1}{2} q_{I}q^{I} - \frac{3}{4} (q_{I} \xi^{I})^2 \right] z^3 \nonumber \\ & + \frac{1}{12}q_{I} \xi^{I} \left[1 + \frac{3}{2}q_{I}q^{I} - \frac{3}{4} (q_{I} \xi^{I})^2 \right] z^4 +  \order{z^5}~,
\end{align}
where $\eta^{IJ}$ is used to raise the indices of $\xi_I$ and $q_I$.
Using this coordinate change and expanding the fields in FG coordinates, the on-shell action and one-point functions are
\begin{align}
I_{\rm ren} &= \text{Vol}(\AdS_2) \text{Vol}(S^1) r_+~, \nonumber \\
\ev{T_{ij}} &= -\frac{q_{I} \xi^{I}}{2} \mqty( -g_{\AdS_2} & 0 \\ 0 & 2g_{S^{1}} )_{ij}~, \nonumber \\
\ev{T^i_i} &= 0 ~, \nonumber \\
\ev{\cO_{\alpha}} &= 0~, \nonumber \\
\ev{{J_I}_i} &= \frac{q_I}{\sqrt{2}} \delta_{i\theta}~. 
\end{align}

\section{Regularity}
\label{sec:conical}

In this section, we impose two regularity conditions on the solutions. 
First, we demand that the geometry smoothly closes off at the largest positive zero of $f(r)$ without a conical singularity in the bulk spacetime. This condition is analogous to the regularity condition imposed on Euclidean black hole solutions.  Second, we fix the periodicity of the $S^1$ at the conformal boundary such that when the $\AdS_2\times S^1$ boundary is conformally mapped to $\mathbb{R}^{1,2}$ there is no conical deficit on the boundary. This condition is different from the one imposed in the holographic calculation of supersymmetric R\'enyi entropies \cite{Nishioka:2013haa,Nishioka:2014mwa,Huang:2014pda,Crossley:2014oea},
which use solutions that are related by double analytic continuation. For these solutions, the periodicity is related to the R\'enyi index $n$.

The regularity  conditions will impose constraints on the parameters of the solutions. Since the general solution is only implicit,  a detailed analysis is performed for the examples presented in this paper.
We will show that for the single scalar and coset models, these conditions imply a bound on the expectation value of the boundary stress tensor.

\subsection{General Statements}
\label{sec:conical-general}

Given the metric
\begin{align}
\dd{s^2} = r^2 \sqrt{H(r)} \dd{s^2_{\AdS_2}} + \frac{f(r)}{\sqrt{H(r)}} \dd{s^2_{S^1}} + \frac{\sqrt{H(r)}}{f(r)} \dd{r^2}~,
\end{align}
the center of the space $r = r_+$ is defined to be the largest zero of $f(r) = -1 + r^2 H(r)$.
We can identify four criteria a regular geometry should satisfy:
\begin{itemize}
\item[(a)] positivity of the zero, $r_+ > 0$,
\item[(b)] $0 < H(r) < \infty$ on $r \in [r_+, \infty)$,
\item[(c)] $0 < f(r) < \infty$ on $r \in (r_+, \infty)$, and
\item[(d)] no conical singularity at $r = r_+$.
\end{itemize}
Criteria (b) and (c) are satisfied if $H(r)$ is continuous: the $\AdS$ length scale \eqref{eq:ads-length} is well-defined if and only if the limit $H(r = \infty)$ is positive and finite.
Since a zero of $H(r)$ occurs at $f(r) < 0$, positivity of $H(r)$ at large $r$ and continuity imply that the spacetime closes off before a zero of $H(r)$ is ever encountered. 

By expanding the metric around the center of the space, criterion (d) is satisfied when
\begin{align}
f'(r_+)^2 = 4 H(r_+)~.
\end{align}
This can be simplified to
\begin{align}
H'(r_+) ( r_+^2 f'(r_+)  + 2 r_+) = 0 ~.
\end{align}
As the second factor is the sum of two positive quantities, a conical singularity can be avoided if we satisfy the condition $H'(r_+) = 0$.
As $r_+$ is determined implicitly in terms of the $q_I, \xi_I$ constants through the equation $f(r_+) = 0$, this condition can be viewed as a constraint on the possible values $q_I, \xi_I$ can take.
Additionally, we will see that criterion (a) manifests as an inequality on $q_I, \xi_I$ that we must satisfy.

\subsection{Single Scalar Model}
\label{sec:conical-single}

\begin{figure}[ht]
	\centering
	\begin{subfigure}{.35\textwidth}
		\includegraphics[width=\textwidth]{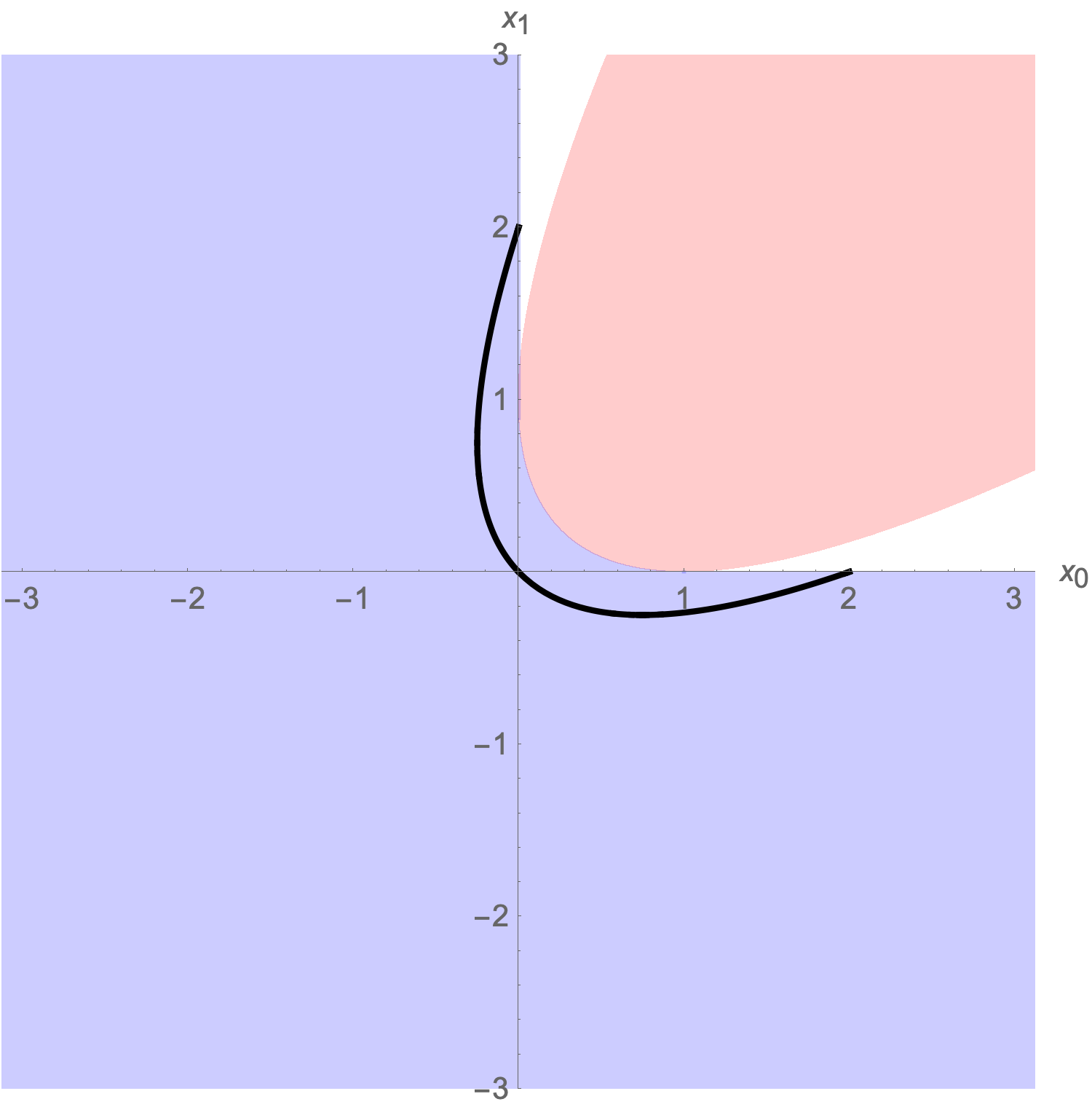}
		\caption{$r_1$}
		\label{fig:r1}
	\end{subfigure}
	\hspace{2cm}
	\begin{subfigure}{.35\textwidth}
		\includegraphics[width=\textwidth]{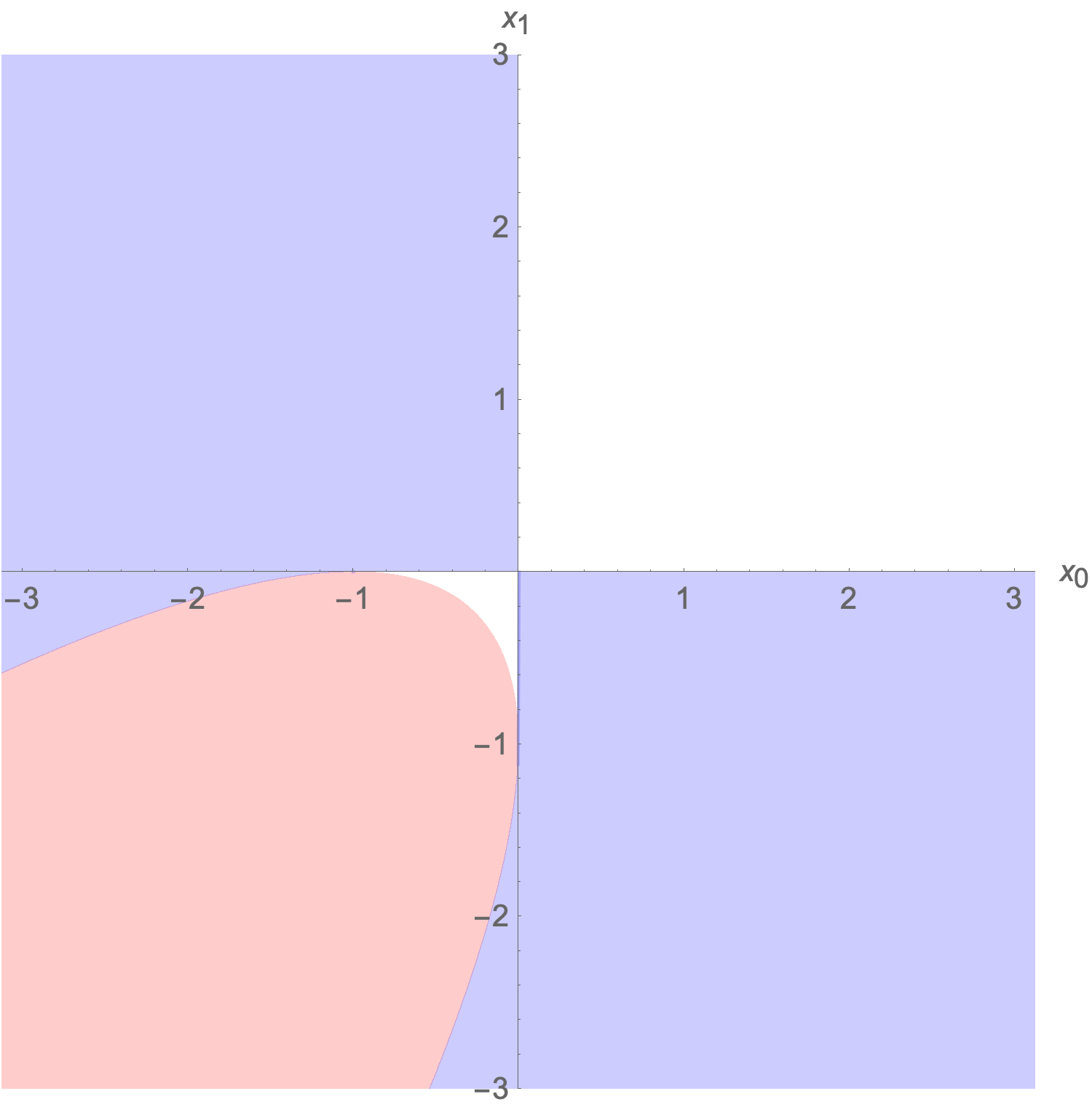}
		\caption{$r_2$}
		\label{fig:r2}
	\end{subfigure}
	\caption{Candidate $r_+$ for the single scalar model.}
\end{figure}

The single scalar model is simple enough that the conditions for a regular geometry can be solved exactly.
Let us define $x_I \equiv q_I/\xi_I$, but still pick the $\AdS$ length scale to be unity, i.e.~keep $\xi_0 \xi_1 = 2$. The metric functions become
\begin{align}
H(r) &= \qty(1 + \frac{x_0}{r})^2 \qty(1 + \frac{x_1}{r})^2~, \nonumber \\
f(r) &= -1 + \frac{1}{r^2} (r+x_0)^2(r+x_1)^2  \label{eq:conical-single}~.
\end{align}
Let us first satisfy the criterion $r_+ > 0$.
Solving $f(r) = 0$,
\begin{align}
0 &= \qty( r^2 + r(x_0 + x_1 - 1) + x_0 x_1 )\qty( r^2 + r(x_0 + x_1 + 1) + x_0 x_1 )~. \label{eq:conical-fzeros}
\end{align}
When the first factor is zero, we have a solution
\begin{align}
r_1 = \frac{1}{2} \qty( -(x_0 + x_1 - 1) + \sqrt{(x_0 + x_1 - 1)^2 - 4 x_0 x_1} )~, \label{eq:conical-r1}
\end{align}
where we took the $+$ sign to get the largest root.
This solution exists when $(x_0 + x_1 - 1)^2 - 4 x_0 x_1 \geq 0$, which is a region on the $x_0x_1$-plane bounded by a parabola, shown in figure \ref{fig:r1}.
The red shaded region indicates where $\smash{r_1}$ does not exist and the blue shaded region indicates where $\smash{r_1 > 0}$. 
When the second factor of \eqref{eq:conical-fzeros} is zero, we have another solution
\begin{align}
r_2 = \frac{1}{2} \qty( -(x_0 + x_1 + 1) + \sqrt{(x_0 + x_1 + 1)^2 - 4 x_0 x_1} )~,
\end{align}
where we also took the $+$ sign.
We have also marked regions where this solution exists and is positive in figure \ref{fig:r2}.
In regions where $\smash{r_1}$ and $\smash{r_2}$ both exist and $r_1 > 0$, we have $\smash{r_1 > r_2}$.
Therefore, we can take $r_+ = \smash{r_1}$ and restrict the $(x_0, x_1)$ parameter space to the blue shaded region of figure \ref{fig:r1}. \\

Let us now avoid the conical singularity by satisfying $H'(r_+) = 0$.
Calculating the derivative of $H(r)$ in \eqref{eq:conical-single} and plugging in $r_+ = r_1$ from \eqref{eq:conical-r1}, we get the condition
\begin{align}
0 = (x_0 - x_1)^2 - 2(x_0+x_1)~. \label{eq:conical-cond}
\end{align}
This is a parabola, marked by the black curve in figure \ref{fig:r1} in the region where $r_+ > 0$.
For the single scalar model to admit a regular geometry, the parameters $x_I = q_I / \xi_I$ must satisfy this condition.
As a corollary, we can note that
\begin{align}
0 \leq x_0 + x_1 < 2~. \label{eq:conical-bound}
\end{align}
This implies that the components of the boundary stress tensor \eqref{eq:single-evs} have bounded expectation value.
Additionally, the pure $\AdS_4$ vacuum ($x_0 = x_1 = 0$) is the only solution with regular geometry and $\ev{T_{ij}} = 0$.

\subsection{$\SU(1, n)$ Coset Model}
\label{sec:conical-coset}

The coset model is also simple enough that the conditions for a regular geometry can be solved exactly.
We can note that 
\begin{align}
H(r) = \qty( 1 - \frac{q_I \xi^I}{r} - \frac{q_I q^I}{2 r^2} )^2 ~,
\end{align}
actually has the same form as \eqref{eq:conical-single}, where
\begin{align}
x_0 &= \frac{- q_I \xi^I - \sqrt{ (q_I \xi^I)^2 + 2 q_I q^I}}{2}~, & x_1 &= \frac{- q_I \xi^I + \sqrt{ (q_I \xi^I)^2 + 2 q_I q^I}}{2}~.
\end{align}
This map is always well-defined as $(q_I \xi^I)^2 + 2 q_I q^I \geq 0$, which can be checked by rotating to the frame where $\xi_I = (\sqrt{2}, 0, 0, \dotsc)$.
Thus all our results for the single scalar model can be carried over.
The bound \eqref{eq:conical-bound} for the single scalar model translates to the same bound on $\ev{T_{ij}}$ for the coset model:
\begin{align}
0 \leq - q_I \xi^I < 2~.
\end{align} 
The condition \eqref{eq:conical-cond} for a regular geometry translates to
\begin{align}
0 = (q_I \xi^I)^2 + 2 q_I q^I + 2 q_I \xi^I~.
\end{align}
We can show that the only regular geometry with vanishing $\ev{T_{ij}}$ is the $\AdS_4$ vacuum.
If we rotate to the frame where $\xi_I = (\sqrt{2}, 0, 0, \dotsc)$, the only $q$ which satisfies $ q_I \xi^I = 0$ and $q_I q^I= 0$ is $q_I = 0$.
A general $\xi$ then has a $q$ in the orbit of $q_I = 0$, which is still the zero vector.

\subsection{Gauged STU Model}
\label{sec:conical-stu}

For the gauged STU model, it is not practical to solve $f(r) = 0$ to find $r_+$ as $f$ is a quartic polynomial.
However, we still expect the criterion $r_+ > 0$ to impose an inequality on the four-dimensional parameter space $(x_0, x_1, x_2, x_3)$ and the condition of avoiding a conical singularity to reduce this to a three-dimensional hypersurface.
However, note that unlike the single scalar and coset models, the expectation value $\ev{T_{ij}}$ is not bounded.
In appendix \ref{sec:stu-conical} we give special cases of the STU model with regular geometry which can have arbitrarily large $x_0 + x_1 + x_2 + x_3$.

\section{Discussion}
\label{sec:discussion}

In this paper, we constructed solutions of four-dimensional $N=2$ gauged supergravity by a double analytic continuation of the half-BPS black hole solutions first found by  Sabra \cite{Sabra:1999ux}. While the black hole solutions exist  for arbitrary prepotentials,  explicit expressions for the scalars fields  involve  algebraic equations which in general can only be solved numerically. We considered three explicit examples of matter-coupled gauged supergravities, namely the single scalar model, the gauged STU model, and the  $\SU(1,n)/\U(1)\times \SU(n)$ coset model to find solutions and calculate holographic observables.  

The solutions we find are holographic duals to line defects in three-dimensional SCFTs. 
The defect is characterized by a non-trivial expectation value of the $R$-symmetry and flavor currents along the $S^1$ factor in the $\AdS_2\times S^1$ description of the defect. After conformally mapping to Minkowski space, this corresponds to a holonomy when encircling the line defect.  
The expectation values of the real scalar operators vanish for general models as a consequence of supersymmetry.

For a conformal defect on $\AdS_2\times S^1$, the expectation value of the stress tensor can be parameterized by a single coefficient $h$,
\begin{align}
\ev{T_{ab}}&= h \;g^{\AdS_2}_{ab}~, \quad \ev{T_{\theta\theta} }= -2 h \;g_{\theta\theta}~,
\end{align}
in analogy  to the scaling dimension of local operators \cite{Kapustin:2005py,Drukker:2008jm}.
However,  there are in general no unitarity bounds on  $h$ which follow from the superconformal algebra. For line operators in $N=4$ SYM and ABJM theories, $h$ can be related to the so-called Bremsstrahlung function $B$  \cite{Lewkowycz:2013laa, Fiol:2015spa,Liendo:2016ymz,Bianchi:2017ozk,Bianchi:2018scb} which has been used in the application of conformal booostrap techniques to the study of defects \cite{Gadde:2016fbj,Fukuda:2017cup,Billo:2016cpy,Lauria:2017wav}.
For the single scalar and coset models studied in this paper  we find that $-2<h\leq0$, where the upper bound is saturated only by the $\AdS_4$ vacuum. However,  such a bound does not seem to  generally hold, since  
for the gauged STU model,  $h$ can become arbitrarily negative. Based on numerical searches we conjecture that only the $\AdS_4$ vacuum has vanishing $h$. Note that recently the relation of $h$ and $B$, as well as the negativity of  $h$ has been established on the SCFT side for various defect theories \cite{Bianchi:2018zpb,Lemos:2017vnx,Jensen:2018rxu,Bianchi:2019sxz}
and the arguments should carry over to the defects dual to the solutions studied in the paper.\footnote{We thank Marco Meineri and Lorenzo Bianchi for a useful correspondence regarding these matters.}

The solutions we find are related to supergravity solutions \cite{Hosseini:2019and,Nishioka:2014mwa,Huang:2014pda,Crossley:2014oea} which are holographic duals for a supersymmetric version of R\'enyi entropy first formulated in \cite{Nishioka:2013haa}. We note two differences. First, the solutions we find in Minkowski time signature have real gauge fields, unlike the duals cited above.\footnote{After analytic continuation to Euclidean signature, the gauge fields in both cases are real.}  Second,  we impose the condition that the periodicity of the circle in $\AdS_2\times S^1$ boundary is such that after a conformal map we obtain flat space without a conical singularity. On the other hand,  in the holographic duals to the Super-R\'enyi entropy, the conical singularity is related to the R\'enyi index $n$. We note that in \cite{Hosseini:2019and,Nishioka:2014mwa,Huang:2014pda,Crossley:2014oea} the holographic calculation of the R\'enyi entropy was compared to a localization calculation and agreement was found, and it would be interesting to see whether such a calculation can be performed for the holonomy defects described in this paper.  

Another interesting question is whether more general solutions going beyond the examples discussed in this paper can be found. First, it would be interesting to study (numerical) solutions for more complicated superpotentials.  Second, it would be interesting to see whether one can go beyond the gauged supergravity approximation and find solutions dual to holonomy defects in ten- or eleven-dimensional duals of $N=2$ SCFTs.
Uplifting the solutions found in this paper might prove to be a useful guide in this direction  \cite{Cvetic:1999xp}.

\
\section*{Acknowledgements}
The work of M.~G.~was supported, in part, by the National Science Foundation under grant PHY-19-14412. 
All the authors are grateful to the Mani L.~Bhaumik Institute for Theoretical Physics for support.

\newpage

\appendix

\section{Supersymmetry}
\label{sec:susy}

We use the metric conventions $\eta = (- + + +)$ and $\epsilon_{0123} = - \epsilon^{0123} = 1$. 
The gamma matrices are defined as usual, e.g.
\begin{align}
\{ \gamma_a, \gamma_b\} &= 2 \eta_{ab}~, & \gamma_{ab} &= \frac{1}{2} [ \gamma_a, \gamma_b ]~, & \gamma_5 &= i \gamma_0\gamma_1\gamma_2\gamma_3~.
\end{align}
The two chiral gravitinos can be written in terms of a single complex (Dirac) spinor $\psi_\mu$,  and likewise for the gauginos $\lambda^\alpha$.
The supersymmetry transformations of the four-dimensional gauged supergravity are \cite{Cacciatori:2008ek}
\begin{align}
\delta \psi_\mu =~& \bigg( \partial_\mu + \frac{1}{4} \omega_\mu^{ab} \gamma_{ab} + \frac{i}{2} Q_\mu \gamma_5 + i g \xi_I A^I_\mu + g e^{\cK/2} \gamma_\mu \xi_I \qty( \Im Z^I + i \gamma_5 \Re Z^I ) \nonumber \\
& + \frac{i}{4} e^{\cK/2} \gamma^{ab} (\Im \cN)_{IJ} \qty( \Im(F^{-I}_{ab} Z^J) - i \gamma_5 \Re(F^{-I}_{ab} Z^J) ) \gamma_\mu \bigg) \epsilon~, \nonumber \\
\delta \lambda^\alpha =~& \bigg( \gamma^\mu \partial_\mu \qty( \Re z^\alpha - i \gamma_5 \Im z^\alpha) + 2 g e^{\cK/2} \xi_I  \qty( \Im(  \cD_{\bar\beta} \bar{Z}^I g^{\alpha\bar\beta}) - i \gamma_5 \Re( \cD_{\bar\beta} \bar{Z}^I g^{\alpha\bar\beta} )) \nonumber \\
&+ \frac{i}{2} e^{\cK/2} \gamma^{ab} (\Im \cN)_{IJ} \qty( \Im( F^{-I}_{ab}\cD_{\bar\beta} \bar{Z}^J g^{\alpha\bar\beta} ) - i \gamma_5 \Re( F^{-I}_{ab}\cD_{\bar\beta} \bar{Z}^J g^{\alpha\bar\beta} ) ) \bigg) \epsilon~, \label{eq:bps}
\end{align}
where $\epsilon$ is a complex spinor, and we have defined
\begin{align}
F^{\pm I}_{ab} &\equiv \frac{1}{2} (F^I_{ab} \pm \tilde{F}^I_{ab} )~, & \tilde{F}^I_{ab} &\equiv - \frac{i}{2} \epsilon_{abcd} F^{cd}~.
\end{align}
The K\"ahler connection $Q_\mu$ is 
\begin{align}
Q_\mu = -\frac{i}{2} \qty( \partial_\mu \tau^\alpha \partial_\alpha \cK - \partial_\mu \bar{\tau}^{\bar\alpha} \partial_{\bar\alpha} \cK) ~.
\end{align}
For the gauged STU model defect solution \eqref{eq:defect-stu1}, we can work with the explicit coordinates $(x^0, x^1, x^2, x^3) = (t, \eta, \theta, r)$ and the metric
\begin{align}
\dd{s^2} = r^2 \sqrt{H} \qty( \frac{ -\dd{t^2} + \dd{\eta^2} }{\eta^2} ) + \frac{f}{\sqrt{H}} \dd{\theta^2} + \frac{\sqrt{H}}{f} \dd{r^2}~.
\end{align}
The non-vanishing spin connection 1-forms of the metric are
\begin{align}
\omega^{01} &=  - \frac{\dd{t}}{\eta}~, & \omega^{03} &= \frac{f^{1/2}}{H^{1/4}} \dv{r} ( r H^{1/4} ) \frac{\dd{t}}{\eta}~, \nonumber \\
\omega^{13} &= \frac{f^{1/2}}{H^{1/4}} \dv{r} ( r H^{1/4} ) \frac{\dd{\eta}}{\eta}~,  & \omega^{23} &= \frac{f^{1/2}}{H^{1/4}} \dv{r} \qty( \frac{f^{1/2}}{H^{1/4}} ) \dd{\theta}~.
\end{align}
For the following calculations, we use the parametrization $(Z^0, Z^1, Z^2, Z^3) = (i, iz^2 z^3, iz^1z^3, iz^1z^2)$.
The BPS equations \eqref{eq:bps} simplify to
\begin{align}
0 = \delta \psi_\mu &= \qty( \partial_\mu + \frac{1}{4} \omega_\mu^{ab} \gamma_{ab} + i g \xi_I A^I_\mu + \sqrt{2} g  \gamma_\mu \dv{r} (r H^{1/4} ) - \frac{i}{2} \gamma_{23} \gamma_\mu  \dv{r} (H^{-1/4}) )~, \nonumber \\
0 = \delta \lambda^\alpha &= \dv{z^\alpha}{r} \qty( \frac{f^{1/2}}{H^{1/4}} \gamma_3 + 2\sqrt{2} gr H^{1/4} + \frac{i}{H^{1/4}} \gamma_{23} ) \epsilon~.
\end{align}
The gaugino equation implies the projector
\begin{align}
0 = \qty( 1 + \frac{2\sqrt{2} g r \sqrt{H}}{\sqrt{f}} \gamma_3 - \frac{i}{\sqrt{f}} \gamma_2 ) \epsilon~. \label{eq:projector}
\end{align}
The $\mu = t, \eta, \theta$ components of the gravitino equation then simplify to
\begin{align}
0 &= \qty( \partial_t - \frac{1}{2\eta} \gamma_{01} - \frac{i}{2\eta} \gamma_{023} ) \epsilon~,  \nonumber  \\
0 &= \qty( \partial_\eta - \frac{i}{2\eta} \gamma_{123} ) \epsilon~,  \nonumber  \\
0 &= \qty( \partial_\theta + i \sqrt{2} g \qty( -1 + \frac{1}{\sqrt{2}} \xi_I \mu^I ) ) \epsilon~.
\end{align}
These can be integrated to
\begin{align}
\epsilon = \exp\bigg( -i \sqrt{2} g \theta \qty( -1 + \frac{1}{\sqrt{2}} \xi_I \mu^I ) \bigg) \exp\bigg( \frac{i}{2} \gamma_{123} \ln \eta \bigg) \exp \bigg( \frac{t}{2} (\gamma_{01} + i \gamma_{023} ) \bigg) \tilde{\epsilon}(r)~.
\end{align}
We can see that we need $\xi_I \mu^I \in 2\sqrt{2} \mathbb{Z}$ in order for $\epsilon$ to be anti-periodic under the identification $\theta \sim \theta + \pi / \sqrt{2} g$.
The $\mu = r$ component of the gravitino equation simplifies to
\begin{align}
\qty( \partial_r + \frac{1}{8} \frac{H'}{H} + \frac{f'}{8\sqrt{2} g r \sqrt{H} \sqrt{f}} \gamma_3 ) \epsilon~. \label{eq:radialbps}
\end{align}
The gaugino projector \eqref{eq:projector} and the radial equation \eqref{eq:radialbps} take the form of the equation solved in the appendix of \cite{Romans:1991nq}, by identifying
\begin{align}
x &\equiv \frac{2 \sqrt{2} g r \sqrt{H}}{\sqrt{f}}~, & y &\equiv \frac{-i}{\sqrt{f}}~, \nonumber \\
\Gamma_1 &\equiv \gamma_3~, & \Gamma_2 &\equiv \gamma_2~.
\end{align}
The solution is
\begin{align}
\tilde{\epsilon}(r) = \frac{1}{H^{1/8}} \qty( \sqrt{ \sqrt{f} + 2 \sqrt{2} g r \sqrt{H}} - \gamma_2 \sqrt{ \sqrt{f} - 2 \sqrt{2} g r \sqrt{H}} ) (1 - \gamma_3) \epsilon_0~, 
\end{align}
where $\epsilon_0$ is a constant spinor.

\section{Vanishing of Scalar One-Point Functions from Supersymmetry}
\label{sec:scalar-one-point}

The scalar one-point function is given by
\begin{equation}
\langle \bar{\mathcal{O}}_{\bar{\alpha}} \rangle = \lim_{\epsilon \rightarrow 0} \left[ \frac{1}{\epsilon^2} \left(z g_{\beta \bar{\alpha}} \partial_{z} \tau^{\beta} + \partial_{\bar{\alpha}} \mathcal{W} \right) \right]~.
\end{equation}
The derivative of the superpotential $\mathcal{W}$ simplifies to
\begin{equation}
\begin{split}
\partial_{\bar{\alpha}} \mathcal{W} &= \partial_{\bar{\alpha}} \left( -\sqrt{2} e^{\mathcal{K}/2} \xi_{I} |Z^{I} | \right) \\
& = -\frac{1}{\sqrt{2}} e^{\mathcal{K}/2} \xi_{I} \left( \sqrt{\frac{Z^{I}}{\bar{Z}^{I}}} \partial_{\bar{\alpha}} \bar{Z}^{I} + (\partial_{\bar{\alpha}} \mathcal{K}) |Z^{I}| \right)~,
\end{split}
\end{equation}
where $|Z^{I}|^2 = Z^{I}(\tau) \bar{Z}^{I}(\bar{\tau}) $. For real scalars, we can choose a parameterization such that $\bar{Z}^{I} = Z^{I}$. This implies
\begin{equation}
\partial_{\bar{\alpha}} \mathcal{W} = -\frac{1}{\sqrt{2}} e^{\mathcal{K}/2} \xi_{I} \left( \partial_{\bar{\alpha}} \bar{Z}^{I} + (\partial_{\bar{\alpha}} \mathcal{K}) \bar{Z}^{I} \right) = -\frac{1}{\sqrt{2}} e^{\mathcal{K}/2} \xi_{I} \mathcal{D}_{\bar{\alpha}} \bar{Z}^{I}~,
\end{equation}
so that
\begin{equation}
\langle \bar{\mathcal{O}}_{\bar{\alpha}} \rangle = \lim_{\epsilon \rightarrow 0} \left[ \frac{1}{\epsilon^2} \left(z g_{\beta \bar{\alpha}} \partial_{z} \tau^{\beta}  -\frac{1}{\sqrt{2}} e^{\mathcal{K}/2} \xi_{I} \mathcal{D}_{\bar{\alpha}} \bar{Z}^{I} \right) \bigg|_{z=\epsilon} \right]~.
\end{equation}
The gaugino BPS variation in FG coordinates is
\begin{equation}
\left( z \gamma_{3} \partial_{z} \tau^{\beta} - 2i g e^{\mathcal{K}/2} \xi_{I} g^{\beta \bar{\alpha}} \mathcal{D}_{\bar{\alpha}} \bar{Z}^{I} \gamma_{5} \right) \epsilon +  \mathcal{O}(z^3) \epsilon = 0~,
\end{equation}
since $F_{ab} \sim 1/r^2 \sim \mathcal{O}(z^2)$. At $\mathcal{O}(z^2)$, the BPS equations imply
\begin{equation}
z \partial_{z} \tau^{\beta} = \pm 2i g e^{\mathcal{K}/2} \xi_{I} g^{\beta \bar{\alpha}} \mathcal{D}_{\bar{\alpha}} \bar{Z}^{I}~.
\end{equation}  
Without loss of generality, we can choose the upper sign by sending $g \rightarrow -g$ if necessary. After setting $g^2=1/8$ we have
\begin{equation}
\langle \bar{\mathcal{O}}_{\bar{\alpha}} \rangle = \langle \mathcal{O}_{\alpha} \rangle = 0~.
\end{equation}

\section{STU Model Special Cases}
\label{sec:stu-conical}

Here we give a construction for STU models with regular geometry and arbitrarily large $x_0 + x_1 + x_2 + x_3$.
The approach we took to find these models was different than that of section \ref{sec:conical-single}. 
Instead of solving the condition $f = 0$ and then $H' = 0$, we first solved $H' = 0$ and then $f = 0$.
The benefit is that $H'$ is a lower-degree polynomial and is technically simpler to solve.
The downside is that this generates spurious solutions: it is possible that the $r$ we obtain is not the largest root $r_+$, and $r_+$ does not satisfy the equation $H' = 0$. 
These spurious solutions then need to be removed by hand.

To summarize our findings, consider the following construction:
\begin{enumerate}
\item Let $x_0$ be any positive number.
\item Numerically solve the equation
\begin{align}
27 x_1 (x_0 - x_1)^4 = -16 x_0(x_0 + 3 x_1)^2~. \label{eq:conical-proof-1}
\end{align}
Let $x_1$ be the unique solution satisfying $-x_0/3 < x_1 < 0$.
\item Consider an STU model with unit $\AdS_4$ length scale where 
\begin{align}
x_0 &= \frac{q_0}{\xi_0}~, & x_1 = \frac{q_1}{\xi_1} = \frac{q_2}{\xi_2} = \frac{q_3}{\xi_3}~.
\end{align}
Numerically solve the equation $f(r) = 0$ for $r$,
\begin{align}
(r+x_0)(r+x_1)^3 = r^2~. \label{eq:conical-proof-2}
\end{align}
There exist exactly two solutions: a positive solution greater than $-x_1$, and a negative solution less than $-x_0$.
Let $r_+$ be the positive solution.
\item Check that $H'(r_+) = 0$.
This is guaranteed by the following argument.
Consider $r^* = - 4x_0 x_1/(x_0 + 3 x_1) > 0$ which satisfies $H'(r^*) = 0$.
This also satisfies $f(r^*) = 0$, as plugging $r = r^*$ into \eqref{eq:conical-proof-2} simplifies to \eqref{eq:conical-proof-1}, which is satisfied by construction of $x_1$.
But as the positive solution to $f = 0$ is unique, we must have $r_+ = r^*$.
\end{enumerate}
The steps above give a STU model with regular geometry.
To prove that $x_0 + 3x_1$ is arbitrarily large, we need a better bound than $-x_0/3 < x_1 < 0$.
To satisfy \eqref{eq:conical-proof-1} for large $x_0$, we have
\begin{align}
x_1 \sim - \frac{16}{27x_0} .
\end{align}
Therefore $x_0 + 3x_1 \approx x_0$ for large $x_0$, and can be arbitrarily large.

\newpage
\providecommand{\href}[2]{#2}\begingroup\raggedright\endgroup

\end{document}